\documentclass[usenatbib]{mn2e}  %MN-09-1333-MJ submitted 12/7/09
\usepackage{natbib}  %psfig}
\usepackage{epsfig}

\catcode`\@=11
\def\gta{\ifmmode{\,\mathrel{\mathpalette\@versim>\,}}
    \else{$\,\mathrel{\mathpalette\@versim>}\,$}\fi}
\def\lta{\ifmmode{\,\mathrel{\mathpalette\@versim<\,}}
    \else{$\,\mathrel{\mathpalette\@versim<}\,$}\fi}
\def\@versim#1#2{\lower 2.9truept \vbox{\baselineskip 0pt \lineskip
    0.5truept \ialign{$\m@th#1\hfil##\hfil$\crcr#2\crcr\sim\crcr}}}
\catcode`\@=12  
\renewcommand{\[}{\begin{equation}}
\renewcommand{\]}{\end{equation}}
\def\rms{{\sc rms}}

{\newif\ifnotend
\notendtrue
\def\veclist{ABCDEFGHIJKLMNOPQRSTUVWXYZabcdefghijklmnopqrstuvwxyz.}
\def\top#1#2.{#1}
\def\tail#1#2.{#2.}
\loop\expandafter\xdef\csname v\expandafter\top\veclist\endcsname%
{{\noexpand\bf\expandafter\top\veclist}}
\edef\veclist{\expandafter\tail\veclist}
\if\veclist.\notendfalse\fi\ifnotend\repeat}

\def\df{{\sc df}}

\def\fracj#1#2{{\textstyle{#1\over#2}}}
\def\pa{\partial}
\def\ex#1{\left\langle#1\right\rangle}
\def\Rc{R_{\rm c}}\def\vc{v_{\rm c}}

\def\kms{\,{\rm km}\,{\rm s}^{-1}}

\def\Myr{\,{\rm Myr}}

\def\Gyr{\,{\rm Gyr}}
 
\def\pc{\,{\rm pc}}
\def\kpc{\,{\rm kpc}}

\def\e{{\rm e}}
\def\d{{\rm d}}

\def\feh{\hbox{[Fe/H]}}

\def\afe{[\alpha/\hbox{Fe}]}
\def\dex{\,{\rm dex}}
\def\figref#1{Fig.~\ref{#1}}
\newcommand{\beq}{\begin{equation}}
\newcommand{\eeq}{\end{equation}}

\title[Distribution functions for the Milky Way]
{Distribution functions for the Milky Way}

\author[J. Binney]{James  Binney\thanks{E-mail:
binney@thphys.ox.ac.uk}\\
Rudolf Peierls Centre for Theoretical Physics, Keble Road, Oxford OX1 3NP, UK\\
}

\begin{document}

\date{Draft, July 11, 2009}

\pagerange{\pageref{firstpage}--\pageref{lastpage}} \pubyear{2009}

\maketitle

\label{firstpage}

\begin{abstract}
Analytic distribution functions (\df s) for the Galactic disc are discussed.
The \df s depend on action variables and their predictions for observable
quantities are explored under the assumption that the motion perpendicular to
the Galactic plane is adiabatically invariant during motion within the plane.
A promising family of \df s is defined that has several adjustable
parameters. A standard \df\ is identified by adjusting these parameters to
optimise fits to the stellar density in the column above the Sun, and to the
velocity distribution of nearby stars and stars $\sim1\kpc$ above the Sun.
The optimum parameters imply a radial structure for the disc which is
consistent with photometric studies of the Milky Way and similar galaxies,
and that 20 per cent of the disc's luminosity comes from thick disc.  The
fits suggest that the value of the $V$ component of the Sun's peculiar
velocity should be revised
upwards from $5.2\kms$ to $\sim11\kms$. It is argued that the standard \df\
provides a significantly more reliable way to divide solar-neighbourhood
stars into members of the thin and thick discs than is currently used.
The standard \df\ provides predictions
for surveys of stars observed at any distance from the Sun.  It is
anticipated that
\df s of the type discussed here will provide useful starting points for much
more sophisticated chemo-dynamical models of the Milky Way.

\end{abstract}


\section{Introduction}

A major thread of current research is work directed at understanding the
origin of galaxies. There are excellent prospects of achieving this goal by
combining endeavours in three distinct areas: observations of galaxy
formation taking place at high redshift, numerical simulations of the
gravitational aggregation of dark matter and baryons, and studies of the
Milky Way. The latter field is dominated by a series of major observational
programs that started fifteen years ago with ESA's Hipparcos mission, which
returned parallaxes and proper motions for $\sim10^5$ stars \citep{Perryman}.
Hipparcos established a more secure astrometric reference frame, and the
UCAC2 catalogue \citep{UCAC} uses this frame to give proper motions for
several million stars. These enhancements of our astrometric database have
been matched by the release of major photometric catalogues [DENIS
\citep{DENIS}, 2MASS \citep{2MASS}, SDSS \citep{SDSS}] and the accumulation
of enormous numbers of stellar spectra, starting with the Geneva--Copenhagen
Survey \citep[][hereafter GCS]{Nordstrom04,HolmbergNA} and continuing with
the SDSS, SEGUE \citep{Yanny} and RAVE \citep{Steinmetz} surveys -- on completion the
SEGUE and RAVE surveys will provide $0.25\times10^6$ and $\sim10^6$
low-dispersion spectra, respectively.
These spectra yield good radial velocities and estimates of $\feh$ and $\afe$
with errors of $\sim0.1\dex$.  Two surveys (HERMES and APOGEE) are currently
being prepared that will obtain large numbers of medium-dispersion
spectra from which abundances of significant numbers of elements can be
determined. The era of great Galactic surveys will culminate in ESA's Gaia
mission, which is scheduled for launch in late 2011 and aims to return
photometric and astrometric data for $10^9$ stars and low-dispersion spectra
for $>10^7$ stars.

The Galaxy is an inherently complex object, and the task of interpreting
observations is made yet more difficult by our location within it.
Consequently, the ambitious goals that the community has set itself, of
mapping the Galaxy's dark-matter content and unravelling how it was
assembled, can probably only be attained by mapping observational data onto
sophisticated models. We are developing a modelling strategy that has as its
point of departure analytic approximations to the distribution functions (\df
s) of various components of the Galaxy (McMillan et al.\ in preparation).  In
this paper we present such approximations for the thin and thick discs.  The
paper is organised as follows. Section 2 explains how the \df\ is assembled.
Section 3 compares the \df's predictions for various observables to data. In
particular evidence is presented that the Sun's $V$ velocity is
conventionally underestimated by $\sim6\kms$ and predictions are given for
velocity distributions as a function of distance from the plane.  Evidence is
presented that  the  standard \df\ provides a cleaner division of
solar-neighbourhood stars into members of the thin and thick discs than has
been available hitherto. Section 4
sums up and looks ahead.

\section{Choice of the DF}

Our approach to Galaxy modelling starts from \df s that are analytic
functions of the action integrals ($J_r,J_\theta,J_\phi$) of orbits in an
integrable, axisymmetric Hamiltonian (McMillan et al., in
preparation). The action $J_\phi$ associated with the azimuthal invariance of
the Hamiltonian is simply the $z$ component of angular momentum $L_z$, and the
action denoted $J_\theta$ in Binney \& Tremaine (2008; hereafter BT08), which
quantifies motion perpendicular to the symmetry plane $z=0$, is here
conveniently denoted $J_z$.  We shall be largely concerned with orbits that
have sufficiently large values of $L_z$ that a reasonable approximation to
their dynamics can be obtained by considering the motion parallel to the
plane to proceed regardless of the vertical motion, and the vertical motion
to be affected by the motion in the plane only in as much as the latter
causes the force perpendicular to the plane to vary in time slowly enough for
the vertical motion to be adiabatically invariant -- see BT08 \S3.6.2(b) for
a  justification of this approximation. At any radius $R$ we
define the vertical potential
 \[
\Phi_z(z)\equiv\Phi(R,z)-\Phi(R,0),
\]
 where $\Phi(R,z)$ is the full potential. Motion in $\Phi_z$ has the energy invariant
\[
E_z(z,v_z)\equiv\fracj12v_z^2+\Phi_z(z).
\]
Given a value for $E_z$, the vertical action can be obtained from a
one-dimensional integral
 \[
J_z(E_z)={2\over\pi}\int_0^{z_{\rm max}}\d z\,v_z,
\]
 where $\Phi_z(z_{\rm max})=E_z$.

Motion parallel to the Galactic plane is assumed to be governed by the radial
potential
 \[
\Phi_R(R)\equiv\Phi(R,0),
\]
 so the radial action is
 \[
J_r(E_R,L_z)={1\over\pi}\int_{R_{\rm p}}^{R_{\rm a}}\d r\,v_R,
\]
 where $R_{\rm p}$ and $R_{\rm a}$ are the peri- and apo-centric radii and
$v_R\equiv\sqrt{2(E_R-\Phi_R)-L_z^2/R^2}$.

Given a point in phase space, we can evaluate $L_z=J_\phi$, $E_z$ and $E_R$ and thus
obtain $J_r$ and $J_z$, so a given \df\ can  be evaluated at any point
in phase space.

We start from the simplest plausible \df s, which have the form
 \[\label{basicform}
f(J_r,J_z,L_z)=f_1(L_z)f_r(J_r,L_z)f_z(J_z,L_z).
\]
 Here $f_1$ is primarily responsible for determining the surface density of
the disc, $f_r$ controls the degree of epicyclic motion within the disc, and
$f_z$ controls the disc's vertical structure. Since there is a close
relation between a star's angular momentum $L_z$ and the radii to which it
contributes to observables, the appearance of $L_z$ in $f_r$ and $f_z$
makes it possible for the disc to become hotter and/or thinner at small
radii.

\begin{figure}
\centerline{\epsfig{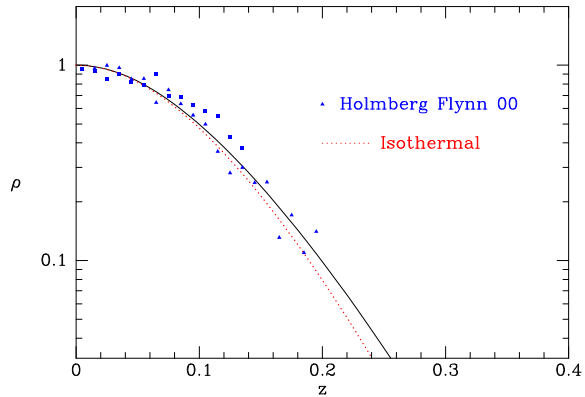}}
 \caption{Comparison of the predictions of two theoretical models and the
density of A (triangles) and F (squares) stars versus distance from the plane
from Holmberg \& Flynn (2000). The full curve is obtained from the \df\
(\ref{basicvert}) with $\sigma_z=6.3\kms$ while the dotted curve is the
classical
exponential $\rho\propto\exp(-\Phi/\sigma_z^2)$ for the same velocity
dispersion. The gravitational potential is that of  Model II of
\S2.7 in BT08.} \label{fig:HF}
\end{figure}

\subsection{Vertical profiles}\label{sec:vert}

We now consider the form of the function $f_z$ in equation
(\ref{basicform}), which controls the disc's vertical structure. We focus on
motion in the solar cylinder of
stars for which $L_z\simeq R_0\vc(R_0)$ so they do not make large radial
excursions. For these stars $f_z$ is effectively a function of only $J_z$. The
classical choice of \df\ is that of an isothermal sheet
$f_z\sim\e^{-E_z/\sigma_z^2}$ \citep{Spitzer42}. Our modelling strategy
requires that we eliminate $E_z$ in favour of $J_z$. In a separable
potential $\Omega_z J_z=\ex{v_z^2}$, where $\Omega_z=\pa
E_z/\pa J_z$ is the vertical frequency and the angle brackets denote a
time average along the orbit with action $J_z$. Moreover, by the virial
theorem $\ex{v_z^2}=E_z$ in a harmonic oscillator, so we replace $E_z$ with
$\Omega_z J_z$ and arrive at what we shall refer to as the ``pseudo-isothermal'' \df
 \[\label{basicvert}
f_{\sigma_z}(J_z)\equiv{\e^{-\Omega_z J_z/\sigma_z^2}
\over2\pi\int_0^\infty\d J_z\,\e^{-\Omega_z J_z/\sigma_z^2}},
\]
 where the denominator ensures that $f_{\sigma_z}$ satisfies the normalisation
condition
 \[\label{normf}
\int\d z\,\d v_zf_{\sigma_z}=1\quad\Leftrightarrow\quad 
\int\d J_z\,f_{\sigma_z}={1\over2\pi}.
\]
 In general $\sigma_z$ is a function of $L_z$ to control the scale height as
a function of radius, but for the moment we neglect this dependence and
investigate the vertical density profile predicted by the \df\
(\ref{basicvert}) by taking $\Phi_z(z)$ to be the potential above the Sun in
Model II of \S2.7 in BT08; this model is disc dominated.

The full curve in \figref{fig:HF} shows the density profile predicted by the
\df\ (\ref{basicvert}) for $\sigma_z=6.3\kms$, while the dotted curve shows the
classical isothermal $\rho\propto\exp(-\Phi/\sigma_z^2)$. The two curves are very
similar because equation (\ref{basicvert}) predicts that $\ex{v_z^2}^{1/2}$
moves in a narrow range from $6.55\kms$ at $z=0$ to a peak value $6.7\kms$ at
$z=240\pc$. Both predictions are in reasonable agreement with the densities
of A and F stars measured by \cite{HolmbergF} shown by triangles and squares,
respectively.

\begin{figure}
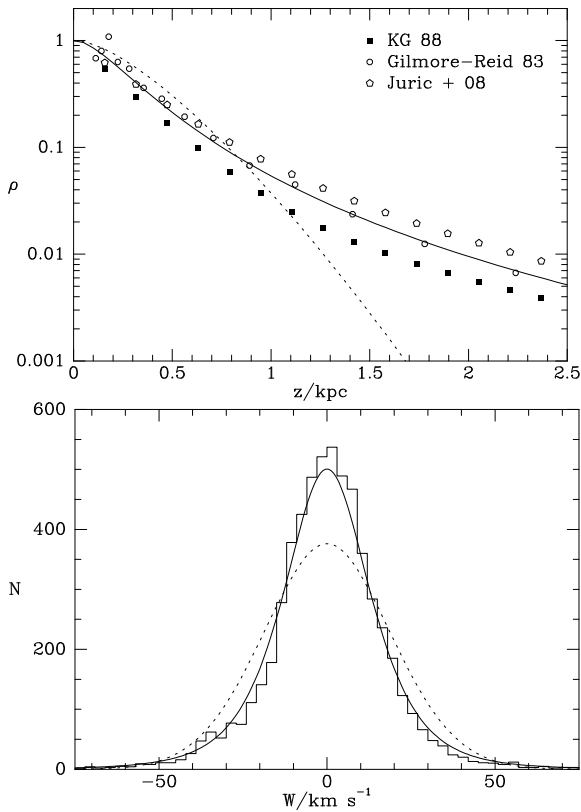

\centerline{\epsfig{file=basicv1.ps,width=.9\hsize}}
\centerline{\epsfig{file=basicv0.ps,width=.9\hsize}}
 \caption{Upper panel: the full curve shows the vertical density profile
predicted in the potential above the Sun by the \df\ of equation
(\ref{powerdf}) with $\gamma=2.6$ and $V_\gamma=18.7\kms$.  The dotted curve
is the profile predicted by equation (\ref{basicvert}) with
$\sigma_z=18\kms$. The points show the density of main-sequence stars
measured by Gilmore \& Reid (1983), Kuijken \& Gilmore (1989) and Juric et
al.\ (2008).  Lower panel: the full and dashed curves show the distributions
in $v_z$ at $z=0$ predicted by the \df s of equations
 (\ref{powerdf}) and (\ref{basicvert}), while the histogram shows the
distribution in $v_z$ of the GCS stars.} \label{fig:vertprof}
\end{figure}

 The dotted curve in the upper panel of \figref{fig:vertprof} shows the
vertical density profile predicted by equation (\ref{basicvert}) when
$\sigma_z=18\kms$. The curvature of the profile has the wrong sign to fit the
density profile of dwarfs with $4<M_V<5$ measured by \cite{GilmoreR}, which
is shown by circles. The velocity dispersion in this model is $\langle
v_z\rangle^{1/2}=19.2\pm0.3\kms$ independent of $z$, so the \df\ is very
close to an isothermal. The dotted curve in the lower panel shows that the
Gaussian distribution in $v_z$ at $z=0$ predicted by the \df\ is a poor fit
to the distribution in $v_z$ of the GCS stars.

One way to obtain a vertical profile that is steeper at small heights and
flatter at large heights is to replace the exponential in equation
(\ref{basicvert}) with an algebraic function of $J_z$. The full curves in
\figref{fig:vertprof} show the density profile and velocity distribution
predicted by the \df
 \[\label{powerdf}
f_z(J_z)={(\Omega_z J_z+V_\gamma^2)^{-\gamma}
\over2\pi\int_0^\infty\d J_z\,(\Omega_z J_z+V_\gamma^2)^{-\gamma}}
\]
 with $\gamma=2.6$ and $V_\gamma=18.7\kms$. The agreement with the data of
\cite{GilmoreR} is
essentially perfect. This fit, and all subsequent fits, were obtained by
adjusting the parameters by hand and judging the quality of the fit by eye.

\begin{figure}
\centerline{\epsfig{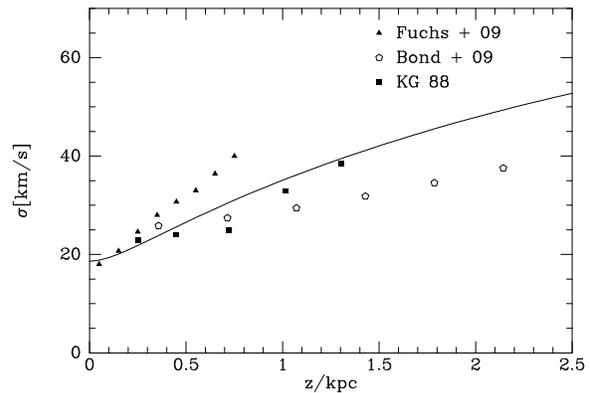}}
\caption{Velocity dispersion as a function of distance from the plane as
predicted by the \df\ (\ref{powerdf}).}\label{fig:powersig}
\end{figure}

The curve in \figref{fig:powersig} shows the extent of the increase in
velocity dispersion with height that is required to produce a thick-disc like
flattening in the density profile at $z>500\pc$: $\ex{v_z^2}^{1/2}$ rises
from $19.6\kms$ at $z=0$ to $37.5\kms$ at $z=1\kpc$ and $50.4\kms$ at
$z=2\kpc$. Also shown are three sets of data points: filled squares show the
values of $\sigma_z$ inferred by \cite{KG89} (hereafter KG89) for the
velocity dispersion of K dwarfs; filled triangles show the dispersion of
metal-poor M dwarfs inferred by \cite{Fuchs09}; open pentagons show the
analytic fit to the dispersions of disc stars that was published by
\cite{Bond09}. Sadly, the data do not tell a coherent story. Each
data set is for a different stellar population, so in principle
their different trends in $\sigma_z(z)$ could be matched by different density
profiles. Therefore in \figref{fig:vertprof} we plot the density profiles
associated with the Bond et al.~ sample \citep[from][]{Juric08} and the KG89
sample. We see that the Juric et al.\ density profile is less steep than the
one from KG89, which is inconsistent with the higher velocity dispersions
measured by KG89. \cite{Fuchs09} do not give a density profile, but it would
be surprising if the kinematics and vertical structure of the M dwarfs were
fundamentally different from that of the K dwarfs studied by KG89, so the
extremely rapid increase in their values of $\sigma_z(z)$ shown in
\figref{fig:powersig} is hard to understand.

The fact that the model curve in \figref{fig:powersig} agrees best with the
KG89 may reflect the fact that the potential in which the \df\ is evaluated
was constrained to be compatible with value for the surface density of
material that lies within $1.1\kpc$ of the plane given by \cite{KG91},
which was based on the KG89 data. In a gravitational potential tailored for
the data of \cite{Juric08} and \cite{Bond09} the \df\ might reproduce the
data in these papers better than that of KG89. We do not pursue this question
here.

Despite the simplicity of the \df\ (\ref{powerdf}) and the accuracy with
which it fits the data, we will not employ it further because it does not
provide a decomposition of the disc into populations of different ages, and
there is no natural way of incorporating it into a \df\ that  also describes
the radial structure of the disc.

Star formation is known to have continued in the disc throughout the life of
the Galaxy and the velocity dispersion of any cohort of coeval stars is known
to increase secularly as a result of scattering by spiral arms and molecular
clouds \citep{SpitzerS,CarlbergS}.  Moreover, as the Galaxy ages, the
chemical composition of the stars that are forming at a given radius changes,
so stars formed at different times and different radii are in principle
distinguishable. Hence it is useful to consider the disc's aggregate \df\ to
be a sum of the \df s for stars of different ages and velocity dispersions.

\begin{figure}
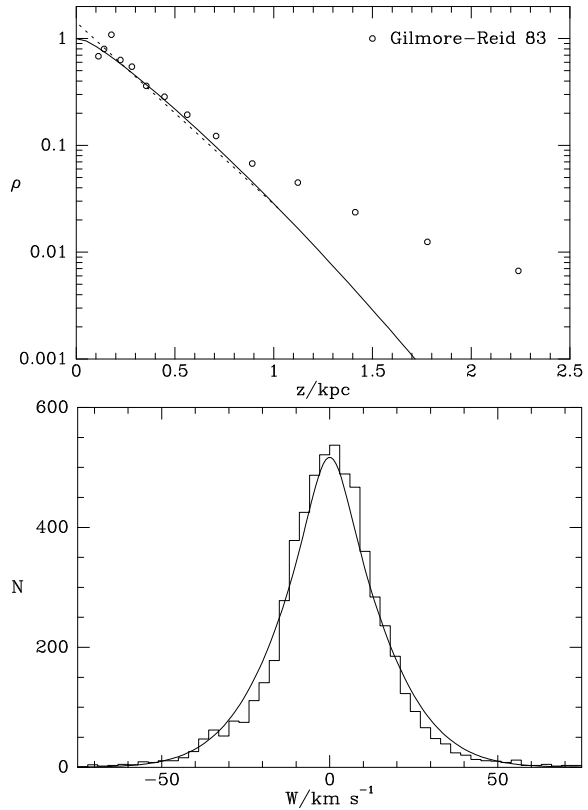

\centerline{\epsfig{file=thin1.ps,width=.9\hsize}}
\centerline{\epsfig{file=thin0.ps,width=.9\hsize}}
 \caption{Full curves: the vertical density profile (upper panel) and the
distribution of $v_z$ (lower panel) predicted by the composite \df\ (\ref{compoDF}). Dashed
line: an exponential of scale height $255\pc$.}\label{fig:thinD}
\end{figure}

We assume that the \df\ of stars of age $\tau$ is the ``pseudo-isothermal''
\df\ (\ref{basicvert}) with $\sigma_z$ increasing with $\tau$ according to
\citep[e.g.][]{AumerB}
 \[\label{sigmatau}
\sigma_z(\tau)=\sigma_{z0}\left({\tau+\tau_1\over\tau_{\rm m}+\tau_1}\right)^\beta.
\]
 Here $\sigma_{z0}$ is the velocity dispersion of stars at age $\tau_{\rm
m}\simeq10\Gyr$, $\tau_1$ sets velocity dispersion at birth, and
$\beta\simeq0.38$ is an index that determines how $\sigma_z$ grows with age.
If we further assume that the rate of star formation has declined with time
as $\e^{-t/t_0}$, then the aggregate \df\ will be
 \[\label{compoDF}
{f}_{\rm thn}(J_z)={\int_0^{\tau_{\rm
m}}\!\!\d\tau\,\e^{\tau/t_0}f_{\sigma_z}(J_z)
\over t_0(\e^{\tau_{\rm m}/t_0}-1)},
\]
 where $\sigma_z$ depends on $\tau$ through equation (\ref{sigmatau}).  In
\figref{fig:thinD} we plot the vertical density profile produced by this
aggregate \df\ when $t_0=8\Gyr$, $\tau_1=0.1\Gyr$, $\sigma_{z0}=20\kms$
\citep{AumerB}. We see that with these parameters we obtain a reasonable fit
to the thin disc.  Specifically, at $z\gta150\pc$ the density is nearly
exponential with a scale height of $255\pc$.  In view of the dramatic
difference between this pure thin-disc structure and the thin plus thick disc
structure furnished by the algebraic \df\ (\ref{powerdf}), it is perhaps
surprising that the velocity distribution in the lower panel of
\figref{fig:thinD} differs as little as it does from the dashed curve in the
lower panel of \figref{fig:vertprof}. This comparison illustrates an
important point: thick-disc stars spend relatively little time near $z=0$ so
they contribute only inconspicuous wings to the velocity distribution there.
The velocity distribution predicted by $f_{\rm thn}$ is close to Gaussian:
the velocity dispersion rises from $16.5\kms$ at $z=0$ to $19.6\kms$ at
$z=500\pc$ and $20.9\kms$ at $z=1\kpc$.

\begin{figure}
\centerline{\epsfig{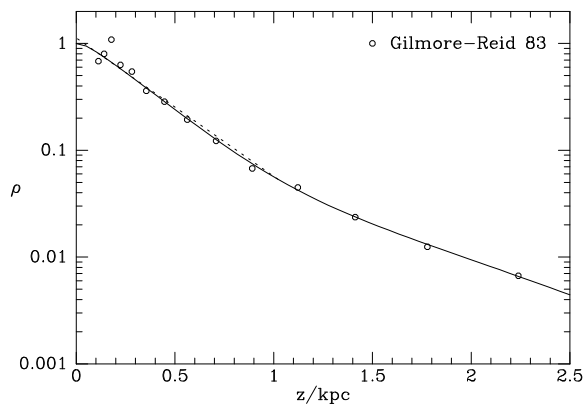}}
\caption{The result of adding to the composite thin disc of
\figref{fig:thinD} a pseudo-isothermal disc with
$\sigma_z=38\kms$ that contains 20 percent of the total mass. The dashed line is an exponential with scale height
$336\pc$.}\label{fig:thickthin}
\end{figure}

\figref{fig:thickthin} shows that a perfect fit to the Gilmore \& Reid (1983)
measurements can be obtained by adding a pseudo-isothermal component with
$\sigma_z=38\kms$  to the thin disc shown in
\figref{fig:thinD}. Within  this
structure $\ex{v_z^2}^{1/2}$ increases from $19\kms$ at $z=0$ to $39\kms$ at
$z=1.5\kpc$ and then very slowly increases to $41.5\kms$ at $z=2.5\kpc$. It
is worth noting that adding the thick disc increases the scale height in the
exponential fit to the profile at low $z$ from $255\pc$ to $336\pc$.  

\subsection{Profiles within the plane}\label{sec:plane}

\cite{Shu69} discussed \df s for planar discs of
the form
 \[\label{ShuDF}
f_r(E,L_z)=\e^{-(E-E_{\rm c})/\sigma_r^2},
\]
 where $E_{\rm c}(L_z)$ is the energy of a circular orbit of angular momentum
$L_z$ and $\sigma_r(L_z)$ is a function that determines the velocity
dispersion in the disc as a function of radius. By analogy with the vertical
\df\ we could replace $E-E_{\rm c}$ by $\Omega_rJ_r$, where $\Omega_r=\pa
E/\pa J_r$ \citep{Binney87,Dehnen99}.  However, the decrease in $\Omega_r$ as
$J_r\to\infty$ is so rapid that for sufficiently eccentric orbits the product
$\Omega_rJ_r$ decreases with increasing $J_r$.  Consequently, if one
substitutes $\Omega_rJ_r$ for $E-E_{\rm c}$, at fixed $L_z$ and large $J_r$
the \df\ increases with $J_r$.  To prevent this unphysical behaviour we
replace $E-E_{\rm c}$ by $\kappa J_r$, where
 \[
\kappa(L_z)\equiv\lim_{J_r\to0}\Omega_r(J_r,L_z)
\]
 is the epicycle frequency. Hence in this paper we adopt as the planar \df\
of a pseudo-isothermal population 
 \[\label{isoJr}
f_r(J_r,L_z)=\e^{-\kappa J_r/\sigma_r^2}.
\]

In view of the normalisation condition (\ref{normf}), the mass $\d M$ placed
by the \df\ (\ref{basicform}) on
orbits with angular momentum in the range $(L_z,L_z+\d L_z)$ is
 \[
\d M=(2\pi)^2f_1\d L_z\int_0^\infty\d J_r\,\e^{-\kappa
J_r/\sigma_r^2}=(2\pi)^2{\sigma_r^2\over\kappa}f_1\d L_z
\]
 In the limit $\sigma_r\to0$ of a cold disc, only circular orbits are populated
and this mass is equal to the mass $2\pi\Sigma R\,\d R$ in the annulus
$(R,R+\d R)$, where $\Sigma$ is the disc's surface density. Hence for a cold disc
 \[
f_1(L_z)={\kappa R\Sigma\over2\pi\sigma_r^2}{\d R\over\d
L_z}={\Omega\Sigma\over\pi\sigma_r^2\kappa}\bigg|_{\Rc},
\]
 where $\Rc(L_z)$ is the radius of the circular orbit of angular
momentum $L_z$ and the second equality uses the identity $\d L_z/\d
R=R\kappa^2/2\Omega$. Here we consider the case of an exponential disc,
$\Sigma=\Sigma_0\e^{-(R-R_0)/R_\d}$, where $R_\d\simeq2.5\kpc$ is the scale
length of the disc and $R_0\simeq8\kpc$ is the Sun's distance from the
Galactic centre. We assume that $\sigma_r$ declines exponentially in
radius with a scale length  that is roughly twice that of the surface density
\[\label{sigofL}
\sigma_r(L_z)=\sigma_{r0}\e^{q(R_0-\Rc)/R_\d}\quad\hbox{where}\quad
q\simeq0.5.
\]
 This choice is motivated by naive epicycle theory, which implies that with
$q\simeq0.5$ the scale height will be constant \citep{vanderKruitS}
provided $\sigma_z/\sigma_r=\hbox{constant}$.

A \df\ such as $f_1$ times equation (\ref{isoJr}) that is an even function of
$L_z$ does not endow the Galaxy with rotation. We introduce rotation by
adding to the \df\ an odd function of $L_z$, which will not contribute to
either the surface density or the radial velocity dispersion. A convenient
choice for this odd contribution to the \df\ is $\tanh(L_z/L_0)$ times the
even contribution, where $L_0$ is a constant that determines the steepness of
the rotation curve in the central region of solid-body rotation.  At radii so
large that $R\vc\gg L_0$ this choice for the odd part of the \df\ simply
eliminates counter-rotating stars. We choose $L_0=10\kms\kpc$, a value
sufficiently small for counter-rotating stars to be confined to the inner
kiloparsec of the Galaxy, which is in reality bulge-dominated.  Hence we
consider the ``pseudo-isothermal'' planar \df
 \[\label{planeDF}
f_{\sigma_r}(J_r,L_z)\equiv{\Omega\Sigma\over\pi\sigma_r^2\kappa}\bigg|_{\Rc}
[1+\tanh(L_z/L_0)]\e^{-\kappa J_r/\sigma_r^2}.
\]

\begin{figure}
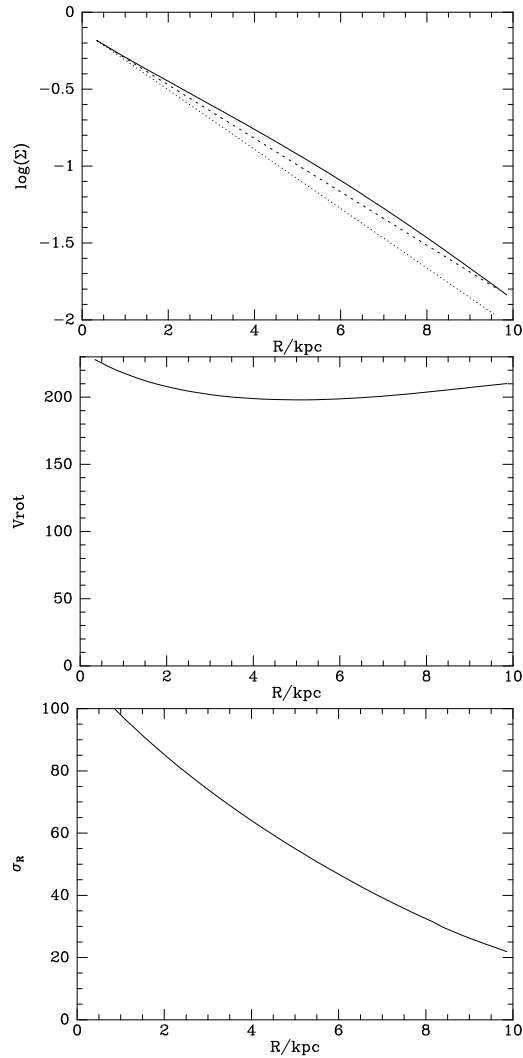

\centerline{\epsfig{file=simple3.ps,width=.8\hsize}}
\centerline{\epsfig{file=simple4.ps,width=.8\hsize}}
\centerline{\epsfig{file=simple5.ps,width=.8\hsize}}
 \caption{From top to bottom, the surface density, rotation speed and radial
velocity dispersion obtained from the \df\ (\ref{planeDF}) when the
gravitational potential is given by equation (\ref{flatPhi}) with
$v_0=220\kms$ and the velocity dispersion function is given by (\ref{sigofL})
with $\sigma_{r0}=28.6\kms$ and $q=0.45$. In the top panel the
dashed line shows an exponential with scale length $2.5\kpc$; 
in equation (\ref{planeDF}) $R_\d=2.25\kpc$ was used  but the
recovered surface density profile is clearly shallower than the exponential
with this scale length, which is shown by the dotted line.
}\label{fig:one}
\end{figure}

It is interesting to evaluate the observables predicted by the \df\
(\ref{planeDF}) when the circular speed is a power law in $R$,
$v_c=v_0(R/R_0)^{\alpha/2}$. Then
 \[\label{powerPhi}
\Phi_R(R)={v_0^2\over\alpha}({R/R_0})^\alpha.
\]
 In the limit $\alpha\to0$ of a perfectly flat circular speed
\[\label{flatPhi}
\Phi_R(R)=v_{\rm 0}\ln(R/R_0).
 \]
 \figref{fig:one} shows the surface density, rotation curve and radial
velocity-dispersion profile predicted by the pseudo-isothermal \df\ (\ref{planeDF})
for a flat circular-speed curve with $v_0=220\kms$. For the plotted profiles
 the function 
$\Sigma(\Rc)$ has been taken to be an exponential of scale length
$2.25\kpc$, while the full curve in the top panel shows that the surface
density produced by this choice of $\Sigma(\Rc)$ provides a good
approximation to the surface density of an exponential disc with a longer
scale length,
$2.5\kpc$, which is shown by the dashed line. At
the price of replacing the analytic function
 \[\label{discSD}
\Sigma(\Rc)=\e^{(R_0-\Rc)/R_\d}
\]
 with a tabulated function, the surface density can be made exactly
exponential \citep{Dehnen99}. Here we adopt the simpler expedient of using a
slightly smaller value of $R_\d$ than the scale length of the disc we wish to
produce.

In the middle panel of \figref{fig:one} the mean rotation speed declines from
$225\kms$ at $R=1\kpc$ to $\sim200\kms$ at $R=5\kpc$ before slowly rising to
$210\kms$ at $R=10\kpc$. The bottom panel shows that the radial velocity
dispersion declines throughout the disc as expected, being $\sim 34\kms$ at
$R_0$.

\begin{figure}
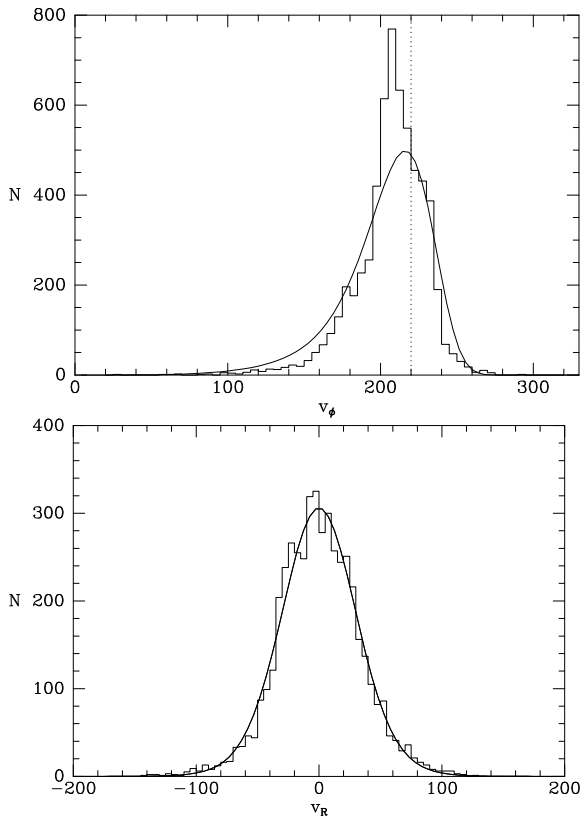

\centerline{\epsfig{file=simple1.ps,width=.9\hsize}}
\centerline{\epsfig{file=simple2.ps,width=.9\hsize}}
 \caption{Comparison of the observed density of GCS stars in the
 $(v_R,v_\phi)$ plane with the prediction of the planar pseudo-isothermal \df\
(\ref{planeDF}) with parameters as in \figref{fig:one}. The smooth curves
show the projections of the model density onto the $v_\phi$ and $v_R$ axes,
while the histograms show the corresponding distributions of stars in the
GCS.}\label{fig:DF}
\end{figure}

\figref{fig:DF} compares the distributions of $v_\phi$ and $v_R$ velocities
predicted by the pseudo-isothermal \df\ with the corresponding distributions of GCS
stars with heliocentric velocities converted to $v_\phi$ and $v_R$ assuming
that the circular speed is $220\kms$ and the Sun's velocity is $v_R=-10\kms$,
$v_\phi=v_{\rm c}+5.2\kms$ \citep[][hereafter DB98]{DehnenB}.  The theoretical and observed
distributions in $v_R$ are satisfyingly similar, but the theoretical $v_\phi$
distribution lies above the observed distribution at small $v_\phi$ and well
below it at $v_\phi\sim200\kms$. We return to this issue below.

As discussed in Section \ref{sec:vert}, realistically we must consider the
thin disc to be a superposition of pseudo-isothermal cohorts of different ages and
chemical compositions. The temperature of the pseudo-isothermal \df\
(\ref{planeDF}) is set by the parameter $\sigma_{r0}$ through equation
(\ref{sigofL}).  By analogy with equation (\ref{sigmatau}), we should
make $\sigma_r$ age dependent by adding to the right side of this equation
the appropriate function of $\tau$. Then $\sigma_r$ is given by
 \[\label{sigofLtau}
\sigma_r(L_z,\tau)=\sigma_{r0}\left({\tau+\tau_1\over\tau_{\rm
m}+\tau_1}\right)^\beta\e^{q(R_0-\Rc)/R_\d}.
\]
 With $\sigma_r(L_z,\tau)$ thus defined, it is
natural to consider the thin-disc \df
 \[\label{thin2d}
\overline{f}(J_r,L_z)={\int_0^{\tau_{\rm m}}\d\tau\,\e^{\tau/t_0}
f_{\sigma_r}(J_r,L_z)\over t_0(\e^{\tau_{\rm m}/t_0}-1)}.
\]
 where $f_{\sigma_r}$ is defined by equation (\ref{planeDF}) with
$\sigma_r(L_z,\tau)$ now obtained from equation (\ref{sigofLtau}).  The \df\
$\overline{f}(J_r,L_z)$ correctly yields values of observables averaged
through the thickness of the disc.  In the case of an external galaxy such
averaged observables are of interest, but samples of Milky-Way stars rarely
if ever provide a sample that is unbiased in $z$. In particular, stars with
large $J_z$ spend little time near the Sun so samples of local stars are
biased against them. Since a star with large $J_z$ is likely to be old,
it is likely to have large $J_r$ also. Hence samples of local stars are
biased towards small $J_r$ and the \df\ (\ref{thin2d}) cannot be used to
predict the properties of solar-neighbourhood stars, or indeed stars at that
lie within any restricted range in $z$.  Instead we must use the full thin-disc
\df.

\section{Full DF}

Putting together the planar and vertical parts of the \df\ for the thin disc
introduced above, we have
 \[\label{thinDF}
f_{\rm thn}(J_r,J_z,L_z)={\int_0^{\tau_{\rm m}}\d\tau\,\e^{\tau/t_0}
f_{\sigma_r}(J_r,L_z)f_{\sigma_z}(J_z)
\over t_0(\e^{\tau_{\rm m}/t_0}-1)},
\]
 where $f_{\sigma_r}$ is defined by equations (\ref{planeDF}) and
(\ref{sigofLtau}), and $f_{\sigma_z}$ is defined by equation
(\ref{basicvert}) but with $\Omega_z$ and $\sigma_z$ now functions of $L_z$
through $\Rc$. By analogy with equation (\ref{sigofLtau}) we have
 \[\label{sigzofLtau}
\sigma_z(L_z,\tau)
=\sigma_{z0}\left({\tau+\tau_1\over\tau_{\rm m}+\tau_1}\right)^\beta
\e^{q(R_0-\Rc)/R_\d}.
\]
 An orbit's vertical frequency $\Omega_z$ is a function of all three actions,
$J_r$, $J_z$ and $L_z$. However, the Jeans theorem assures us that the \df\
remains a solution of the collisionless Boltzmann equation if in $\Omega_z$
we set $J_r=0$. Restricting the
$J_r$-dependence of the \df\ in this way makes the \df\ easier to work with
and is consistent with the work of Section \ref{sec:vert}, which was
restricted to orbits that have $J_r=0$ (``shell orbits''). Therefore in the
following we do this.

\begin{table}
\begin{center}
\caption{Parameters of the standard \df\ (upper section) and values used by
Bensby et al.\ (2003) (lower section)}
\begin{tabular}{lcc}
&Thin disc&Thick disc\\
\hline
$L_0$&$10\kms$&$10\kms$\\
$R_\d$&$2.25\kpc$&$2.3\kpc$\\
$q$&0.45&0.45\\
$\sigma_{r0}$&$33.5\kms$&$60\kms$\\
$\sigma_{z0}$&$19\kms$&$32\kms$\\
$k_{\rm thk}$&-&0.24\\
$\beta$&0.33&-\\
$t_0$&$8\Gyr$&-\\
$\tau_1$&$110\Myr$&-\\
$\tau_{\rm m}$&$10\Gyr$&-\\
\hline
$\sigma_U$&$35\kms$&$67\kms$\\
$\sigma_V$&$20\kms$&$38\kms$\\
$\sigma_W$&$16\kms$&$35\kms$\\
$V_{\rm a}$&$15\kms$&$46\kms$\\
\hline
\end{tabular}\label{tab:params}
\end{center}
\end{table}
 Our final thin-disc \df\ (\ref{thinDF}) is characterised by eight free
parameters, $L_0$, $R_\d$, $q$, $\sigma_{r0}$, $\sigma_{z0}$, $\beta$, $\tau_1$ and
$\tau_{\rm m}$.

\begin{figure}
\centerline{\epsfig{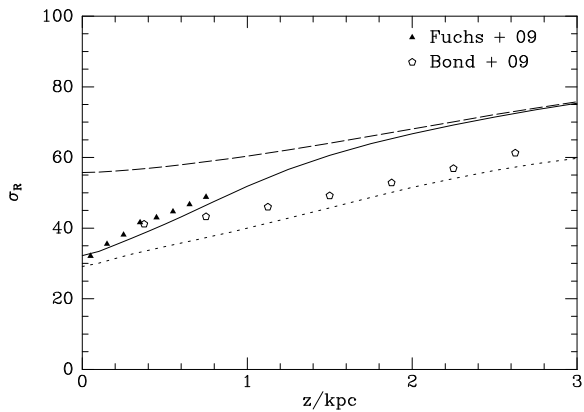}}
\caption{The full curve shows the  radial velocity dispersion in the standard
model. The dotted and dashed  lines show the dispersions in the thin- and
thick-disc components, respectively. The triangles show estimates of $\sigma_R$ for metal-poor M dwarfs
from Fuchs et al.\ (2009), while the open pentagons show the analytic fit to
$\sigma_R(z)$ given by Bond et al.\ (2009)}\label{fig:sigr}
\end{figure}

\begin{figure}
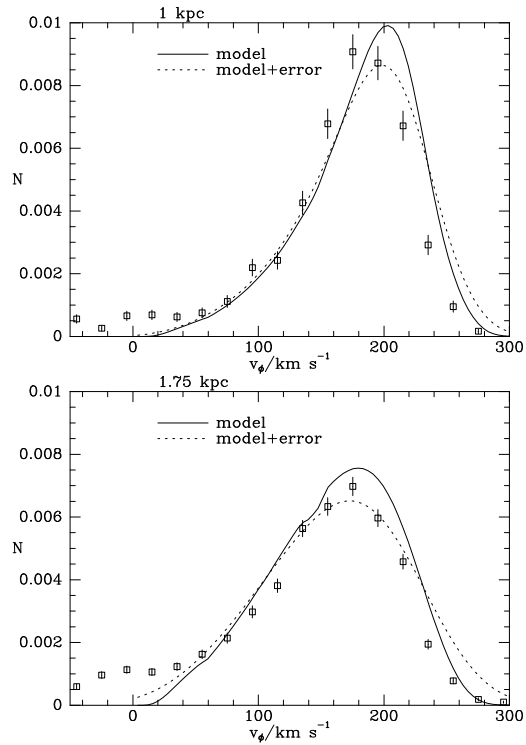

\centerline{\epsfig{file=ivezic0.ps,width=.9\hsize}}
\centerline{\epsfig{file=ivezic1.ps,width=.9\hsize}} \caption{Distributions
of $v_\phi$ velocities $1\kpc$ (top panel) and $1.75\kpc$ (lower panel) above
the plane. The data points are from Ivezic et al.\ (2008) for stars with
$0.8\kpc<z<1.2\kpc$ (upper panel) and $1.5\kpc<z<2\kpc$ (lower panel). The
full curves show the model distributions at $z=1\kpc$ and $1.75\kpc$. The
dashed curves show the result of convolving this distributions with the
measurement errors of Ivezic et al.}\label{fig:ivezic}
\end{figure}

\subsection{Thick disc DF}

At the end of Section \ref{sec:vert} we saw that the observed vertical
density profile at the Sun can be reproduced by adding to the composite \df\
of the thin disc a pseudo-isothermal component with $\sigma_{z0}=38\kms$ that
contains 20 percent of the mass. This result suggests that we add to the \df\
(\ref{thinDF}) of the thin disc the thick-disc \df
 \[\label{thickDF}
f_{\rm thk}(J_r,J_z,L_z)=f_{\sigma_r}(J_r,L_z)f_{\sigma_z}(J_z),
\] 
 where $f_{\sigma_r}$ and $f_{\sigma_z}$ are defined by equations (\ref{planeDF})
and (\ref{basicvert}) above with $\sigma_r$ and $\sigma_z$
given by
 \begin{eqnarray}
\sigma_r(L_z)&=&\sigma_{r0}\,\e^{q(R_0-\Rc)/R_\d}\nonumber\\
\sigma_z(L_z)&=&\sigma_{z0}\,\e^{q(R_0-\Rc)/R_\d},
\end{eqnarray}
 and in equation (\ref{planeDF}) we use (cf eq.~\ref{discSD})
\[
\Sigma=k_{\rm thk}\e^{(R_0-\Rc)/R_\d}.
\]
 Here $k_{\rm thk}$ is the ratio of thick to thin-disc stars in the solar
cylinder. In principle this thick-disc \df\ introduces a further
six parameters: $L_0$, $R_\d$, $q$,  $\sigma_{r0}$, $\sigma_{z0}$ and
$k_{\rm thk}$. Table~\ref{tab:params} lists all the parameters of the
standard \df. We have not explored the option of using a different value of
$L_0$ for each disc because this parameter has a negligible impact on
comparisons with local
data.

The choice of $\sigma_{z0}$ and $k_{\rm thk}$ for the thick disc is
straightforwardly made by the requirement that the vertical density profile
match the data of \cite{GilmoreR}. The choice of $R_\d$, $q$ and
$\sigma_{r0}$ for the thick disc is more problematic. Clearly these
parameters should be constrained by the radial density and kinematics of the
disc at $z>1\kpc$, where the thick disc is dominant. The strongest
constraints are provided by the SDSS.  \cite{Juric08} found the scalelength
of the thick disc to be similar to that of the thin disc.
\cite{Bond09} give an analytic fit to the dependence of $\sigma_R$ on $z$ out to
$z\simeq4\kpc$, and  \cite{Fuchs09} give
several values of $\sigma_R$ at $z\le800\pc$. Finally \cite{Ivezic08}
provide distributions in $v_\phi$ in several ranges of $z$. Figs
\ref{fig:sigr} and \ref{fig:ivezic} show these data.

Although the  data sets are less clearly inconsistent than they are in
\figref{fig:powersig}, the data from Fuchs et al.\ clearly show a
significantly steeper gradient than those from Bond et al. In the model
$\sigma_R(z)$ has a slope intermediate between these values, and agrees with the
data at $z\lta1\kpc$. At greater heights it lies above the Bond et al.\ data,
just as the model curve does in \figref{fig:powersig}.

The model's predictions for the distribution in $v_\phi$ at $z=1\kpc$ and
$z=1.75\kpc$ are shown in \figref{fig:ivezic} together with data points
from \cite{Ivezic08} the heliocentric data of \cite{Ivezic08} have been
converted to galactocentric velocities assuming $v_\phi(\odot)=225.2\kms$,
which arises because in our adopted potential  the circular speed is
$220\kms$ and the peculiar $V$ velocity of the  Sun is $5.2\kms$
\citep{DehnenB}. The full curves are the true model velocity distributions,
and the dotted curves show the result of convolving these distributions with
the errors reported by Ivezic et al., which are $19.5\kms$ at
$z=1\kpc$ and $31.5\kms$ at $z=1.75\kpc$.  

The model curves fall below the data at $v_\phi\la50\kms$ because in this
region halo stars dominate the data points and the model is for the disc
alone. Elsewhere the dotted curves provide a moderate fit to the data points.
The fit in the upper panel would be improved by moving the data points a few
$\!\kms$ to the right, which would be the effect of the upward revision of
the Sun's peculiar velocity advocated below.  The model curves are slightly
too broad. Reducing the parameter $\sigma_r$ in the thick-disc \df\ makes
them narrower, but this change also shifts the model curves still further to
the right, and thus makes the overall fit less good. The distributions can
also be made narrower by either increasing the thick-disc scalelength $R_\d$
or by decreasing $q$. However either change decreases the importance of stars
at apocentre (which have angular momentum $L_z<L_z(\odot)$) relative to those
at pericentre and thus exacerbates the predicted excess of stars at large
$v_\phi$. Extensive experimentation suggests that the fits shown in
\figref{fig:ivezic} cannot be significantly improved upon with a \df\ of the
form (\ref{thickDF}).

Because they are extracted from proper-motion data, the observational
distributions in \figref{fig:ivezic} are sensitive to the photometric
distances employed. A possible resolution of the conflict in the lower panel
of \figref{fig:ivezic} between the model and data is that the distances
employed are slightly too small: using larger distances would increase
heliocentric velocities and thus cause the observational points to move away
from the Sun's assumed velocity, $v_\phi=225.2\kms$. Another possible
resolution of the conflict between data and models in \figref{fig:ivezic} is
the increasing inaccuracy of the assumption of adiabatic invariance of the
vertical motion as random motions become more important. In a future
publication this possibility will be examined with models based on orbital
tori.

\subsection{The standard DF and the solar neighbourhood}

\begin{figure*}
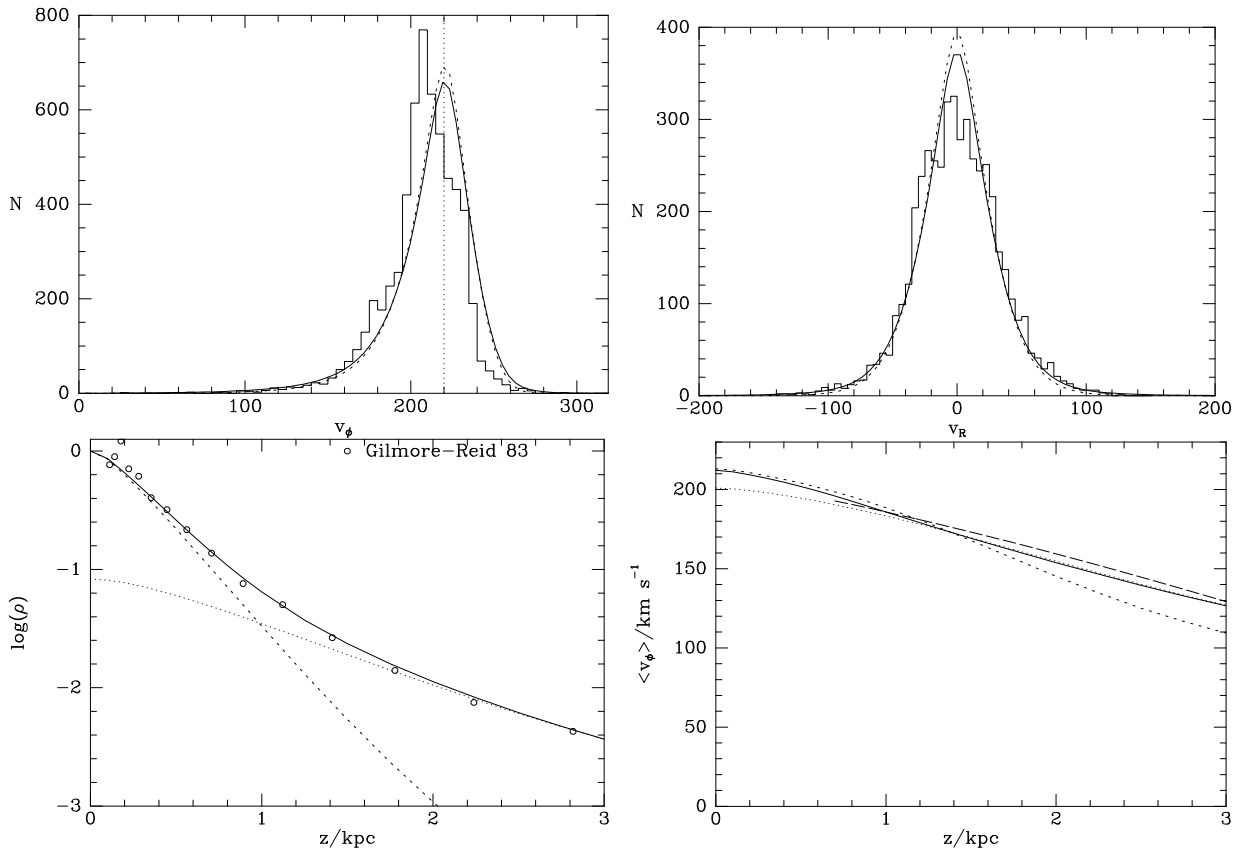

\centerline{\epsfig{file=VphiTT4.ps,width=.45\hsize}\quad
\epsfig{file=VRTT4.ps,width=.45\hsize}}
\centerline{\epsfig{file=rhozTT4.ps,width=.45\hsize}\quad
\epsfig{file=vphizTT4.ps,width=.45\hsize}}
 \caption{Structure at the solar radius predicted by the standard \df\
(eqs.~\ref{thinDF} and \ref{thickDF}).  Full curves are for the entire disc
while dashed curves show the contribution of the thin disc.  The upper panels
show velocity distributions for $z=0$ marginalised over the other velocity
components. The vertical dotted line marks the local circular speed.  In the
lower panels dotted curves show the contributions of the thick disc. In the
bottom-right panel the long-dashed line is the empirical fitting function of
Ivezic et al.\ (2008).  The values of the \df's parameters are given in Table 1.}
\label{fig:thinthick}
\end{figure*}

\figref{fig:thinthick} shows prediction of the standard \df\ for the
structure of the solar neighbourhood. Full curves are for the whole disc and
dashed curves show the contribution of the thin disc. The upper panels show
for stars seen in the plane the distributions in $v_\phi$ and $v_R$ after
integrating over the other two velocity components.  Comparison of these
panels with the panels of \figref{fig:DF} is instructive. The distribution of
$v_R$ velocities is in reasonable agreement with the data, while the
distribution of $v_\phi$ velocities of \figref{fig:thinthick} agrees with the
data better than the distribution in $v_\phi$ in \figref{fig:DF}. Two factors
contribute to the improved fit to the $v_\phi$ velocities. First introducing
a sum of pseudo-isothermals enhances both the core and the wings of the distribution
-- this effect is apparent in the distributions of $v_R$ velocities. More
significantly, including vertical motion suppresses the $v_\phi$ distribution
at low $v_\phi$ because this wing of the distribution is populated by stars
with small values of $\Rc$ that reach the solar neighbourhood because they
have large random velocities. On account of those large velocities, they have
low probabilities of being found close enough to the plane to be included in
the GCS. These stars are most likely to be observed near apocentre, when they
have low values of $v_R$, so depressing the contribution to the GCS of such
stars does not suppress the wings of the distribution of $v_R$ velocities.

The plots shown in \figref{fig:thinthick} are obtained by setting the
function $\Sigma(\Rc)$ for the thin disc that appears in the \df\ to the same
exponential with scale length $2.25\kpc$ that was used to obtain
Figs~\ref{fig:one} and \ref{fig:DF}. The use of $2.25\kpc$ in $\Sigma(\Rc)$
is important not only to ensure that the disc's surface density is
approximately exponential with the larger scale length $2.5\kpc$, but also to
ensure that the predicted distribution of $v_\phi$ velocities agrees with the
GCS data: when $\Sigma$ has scale length $2.5\kpc$, the predicted
distribution falls off too slowly at $v_\phi>\vc$.

The model $v_\phi$ distribution deviates from the data in two respects: it
lacks the pronounced peak in the data centred on $v_\phi=205\kms$, and it
extends too far on the high-velocity wing. The first shortcoming undoubtedly
reflects the axisymmetry of the model and is discussed below. The second
shortcoming, which is also evident in the fits to the data of \cite{Ivezic08}
for the thick disc (\figref{fig:ivezic}), is more interesting. It can be
moderated by increasing the parameter $q$ of equation (\ref{sigofLtau}). This
parameter controls the radial gradient of the thin disc's velocity
dispersion, and a rapid decrease in the amplitude of epicycle motions at
$\Rc>R_0$ limits the number of stars with large angular momentum that can reach
the Sun and thus depopulates the high-$v_\phi$ wing. However, an increase in
$q$ boosts the model profile at low $v_\phi$, so the overall agreement with
the data is not improved unless the parameter $\sigma_{r0}$ of equation
(\ref{sigofLtau}) is simultaneously decreased, and such a decrease leads to
the model $v_R$ distribution being narrower than the data warrant.

Oort's relation e.g., {BT08}  eq.\ (4.317)
 \[
{\sigma_\phi^2\over\sigma_R^2}={-B\over A-B}
\]
 implies that the width of the model $v_\phi$ distribution can be decreased
relative to that of the $v_R$ distribution by changing from a flat to a
falling circular-speed curve.  However, one finds that the relative narrowing
of the $v_\phi$ distribution that is produced by adopting the power-law
potential (\ref{powerPhi}) with $\alpha=-0.2$ produces a negligible improvement on
the fit for constant circular speed shown in \figref{fig:thinthick}. 

The bottom left panel of \figref{fig:thinthick} shows that the overall \df\
provides an excellent fit to the vertical density profile from
\cite{GilmoreR}, and that the vertical profile of the thin disc is extremely
close to exponential. The latter result appears to be fortuitous in that it
involves a subtle interplay between the non-trivial vertical force law and
the number of stars with large random velocities that visit the solar
neighbourhood from significantly nearer the Galactic centre. 

The bottom right panel of \figref{fig:thinthick} shows that in both the thin
and thick discs the mean rotation speed declines with distance from the
plane. In the plane the thin disc rotates faster than the thick disc, as one
naively expects. However, the rotation rate of the thin disc declines faster
with $z$ than that of the thick disc. The slower decline in the thick disc
arises because we have set $R_\d=2.3\kpc$ in the function
$\Sigma(\Rc)=\e^{-\Rc/R_\d}$ for the thick disc -- with $R_\d=2.25\kpc$ for
both discs the thick disc rotates $\sim15\kms$ slower than the thin disc at
all values of $z$. We have chosen $R_\d=2.3\kpc$ for the thick disc to obtain
a better fit to the long-dashed curve in the lower right panel of
\figref{fig:thinthick}, which is an analytic fit to the mean rotation rate
extracted from the proper motions of SDSS stars by \cite{Ivezic08}. The
dependence of the rotation rate on $R_\d$ is consistent with the Stromberg
equation
 \beq\label{eq:Stromberg}
v_{\rm
a}={\sigma_R^2\over2\vc}\left[{\sigma_\phi^2\over\sigma_R^2}-1
-{\partial\ln(\nu\sigma_R^2)\over\partial\ln R}
-{R\over\sigma_R^2}{\partial\sigma_{Rz}^2\over\partial z}\right].
\eeq
 For our preferred \df\ the asymmetric drift of
the thick disc increases from only $20\kms$ at $z=0$ to $90\kms$ at
$z=3\kpc$, consistent with the values $30-50\kms$ usually reported by
observers \citep[e.g.][]{Edvardsson93,GilmoreWK}.

\begin{figure}
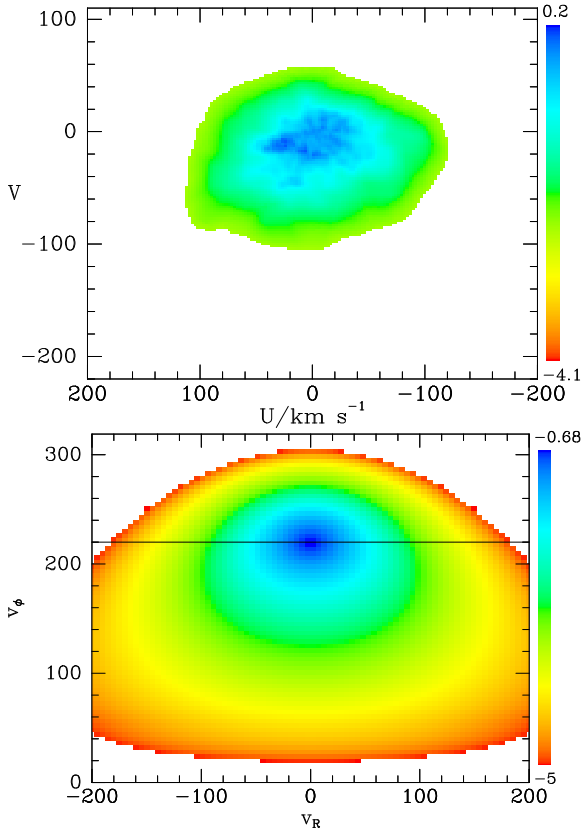

\centerline{\epsfig{file=GCSdens.ps,width=.9\hsize}}
\centerline{\epsfig{file=DFTT4.ps,width=.9\hsize}}
 \caption{The density
of GCS stars in the $(U,V)$ plane (upper panel) and that predicted by the
standard
\df\ (lower panel). Colours indicate $\log_{10}$ of the
stellar density and the horizontal line is at the circular
speed. The upper plot was obtained by applying the FiEstas algorithm of
Ascasibar \& Binney (2005) to  the GCS data and setting the density to zero
if there was no star within $25\kms$ of a point.}\label{fig:UVplane}
\end{figure}

\figref{fig:UVplane} compares the observed density of stars in the plane of
velocities $(U,V)$ with respect to the LSR\footnote{We follow BT08 (p.~12) in
defining the LSR to be $v_R=v_z=0$ and $v_\phi=\vc(R_0)$. The LSR is
sometimes taken to be the speed of a closed orbit through $R_0$. In the
presence of ephemeral spiral structure this second definition is probably not
useful.} (top panel) to that predicted
by the standard \df\ (bottom). The dynamic range in density that can be
sampled with the GCS stars is limited, so only a portion of the standard
\df's predictions for the $(U,V)$ plane is tested. Moreover, the limited
number of GCS stars leads to the steepness of density gradients being
underestimated, for example around $(U,V)=(0,40)\kms$.

Near $(U,V)=(0,0)$, the observational
diagram shows density enhancements, or ``streams'', that are not bounded by
the roughly ellipsoidal surfaces in velocity space on which actions are
constant. Consequently, by Jeans' theorem, the presence of streams indicates
that either the Galaxy's potential is not axisymmetric, or the Galaxy is not
in a steady state -- no \df\ that is a function of actions only can
reproduce these streams, although one hopes to be able to reproduce them by
perturbing such a \df\ using Hamiltonian perturbation theory. The streams
account for much of the disagreement between the theoretical and observed
velocity distributions in \figref{fig:thinthick}. In particular they account
for the peak in the observed $v_\phi$ distribution lying $\sim15\kms$ below
the circular speed.

\subsection{The solar motion} 

We have seen that the \df\ has difficulty simultaneously fitting the
observed distributions in $v_R$ and $v_\phi$ of the GCS stars (top panels of
\figref{fig:thinthick}). A related problem was encountered in the fit to the
proper motions of thick-disc stars (\figref{fig:ivezic}). The agreement
between theory and data in both figures would be improved by shifting the
observed $v_\phi$ distribution to the right. Such a shift would
correspond to increasing the Sun's peculiar velocity by a few kilometers per
second. 

\begin{figure}
\centerline{\epsfig{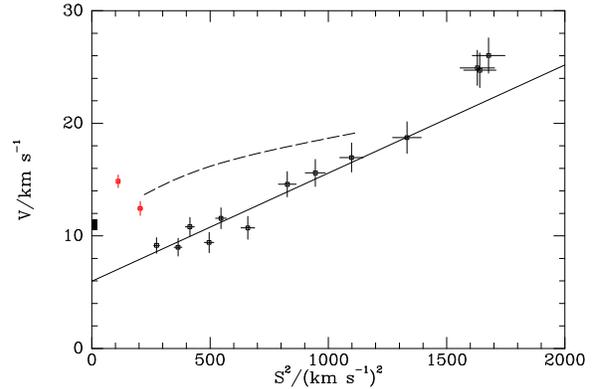}}
 \caption{The data points show, for each group of 1000 main-sequence stars
ordered by colour, the solar motion versus the group's squared
velocity dispersion in the plane of the sky. The points are based on the
proper motions obtained by van Leeuwen (2007) from a reanalysis of the
Hipparcos telemetry for samples of stars defined by Aumer \& Binney (2009).
The straight line is the least-squares fit to the black data points, which
has $y$ intercept at $5.05\kms$. The black rectangle on the $y$ axis shows
the proposed solar motion $V_\odot=11\kms$. The dashed line shows the
solar motions predicted by model \df s for populations ranging in
age from $1$ to $11\Gyr$ under the assumption that $V_\odot=11\kms$ and 
$S^2\simeq(1.2\sigma_R)^2$ (DB98).}\label{fig:solarm}
\end{figure}

Although such a shift significantly exceeds the formal error of $0.6\kms$ on
$V_\odot$ given by DB98, we should not lightly dismiss the possibility that
$V_\odot$ has been underestimated. The value given by DB98 was obtained by
extrapolating to zero velocity dispersion a plot of $V$ velocity versus
squared velocity dispersion $S^2$ for stars grouped by colour such as that
shown in \figref{fig:solarm}. Stromberg's equation (\ref{eq:Stromberg})
suggests that this relation will be linear if the square bracket is constant,
and \figref{fig:solarm} shows that the Hipparcos data are consistent with
this expectation if the groups with the lowest velocity dispersions (shown in
red) are discounted. However, Stromberg's equation is obtained under the
assumption that the Galactic potential is axisymmetric, so in the limit of
vanishing velocity dispersion, stars move on circular orbits. In reality the
potential has a non-axisymmetric component of amplitude $\sim7\kms$, which
manifests itself, inter alia, by causing a plot of terminal velocity versus
Galactic longitude to undulate at this level
\citep[e.g.][Fig.~9.16]{Malhotra,BinneyM}. Given the non-axisymmetric
component of the potential, $\lim_{\sigma\to0}\ex{V}$ can differ from the
velocity of circular motion by of order the amplitude $\sim7\kms$ of the
non-axisymmetric component.

\begin{figure}
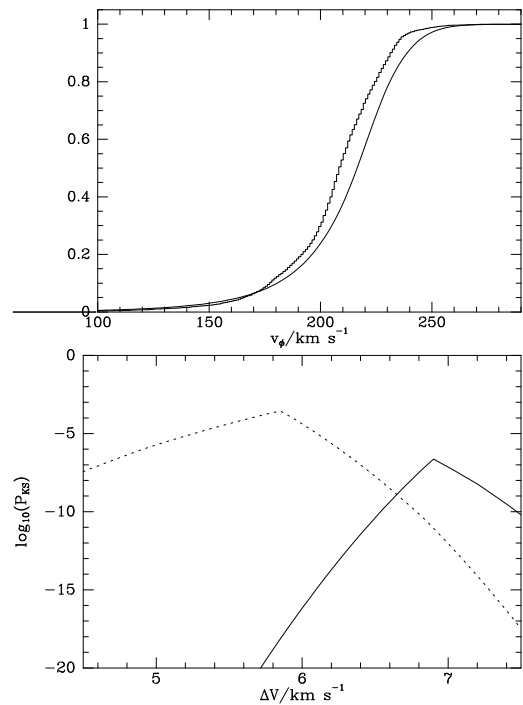

\centerline{\epsfig{file=ksdiff.ps,width=.8\hsize}}
\centerline{\epsfig{file=pks.ps,width=.8\hsize}}
 \caption{Upper panel: the cumulative distributions of stars with $v_\phi$
smaller than the given value for the GCS data (jagged curve) that the
standard \df\ under the assumption  $V_\odot=5.2\kms$ from DB98. Lower panel: the
full curve shows the Kolmogorov--Smirnov probability that the
distribution of $V$ velocities of GCS stars is drawn from a model distribution as a
function of the amount $\Delta V$ added to $5.2\kms$.
Dotted curve: the same but after randomly redistributing those observed stars
that have velocities with respect to the conventional LSR in the range
$(-20,10)\kms$.}
\label{fig:KS}
\end{figure}

As \cite{OllingDehnen} pointed out in their determination of the Oort
constants, the larger the velocity dispersion a population has, the less it
is likely to be affected by non-axisymmetric forces associated with spiral
structure.  The non-axisymmetric potential of the Galaxy's bar is
thought to be responsible for the ``Hercules stream'', an over-density of
stars in velocity space around $v_\phi=\vc-50\kms$, but there is no evidence
that it significantly perturbs the velocity distribution at $v_\phi\gta\vc$,
where the standard model conflicts with the data.    Consequently,
\figref{fig:thinthick} offers an opportunity to determine $V_\odot$ that is
at least as valid as the traditional route using Stromberg's equation.

The jagged curve in the upper panel of \figref{fig:KS} shows the cumulative
distribution in $v_\phi$ of the GCS stars under the assumption that $v_{\rm
c}(R_0)=220\kms$ and $V_\odot=5.2\kms$.  The smooth curve shows the
cumulative distribution of the model shown in \figref{fig:thinthick}. The
need to shift the distribution of GCS stars to the right is evident.

The full curve in the lower panel of \figref{fig:KS} shows the
Kolmogorov--Smirnov probability $P_{\rm KS}$ that the distribution of $V$
velocities of GCS stars is drawn from the model shown in
\figref{fig:thinthick} as a function of the amount $\Delta V$ added to the
solar motion given in DB98.  For all choices of $\Delta V$, $P_{\rm KS}$ is
small, largely due to the impact of streams on the data for $-20\kms\lta
V_{\rm LSR}\lta10\kms$. The impact of streams on $P_{\rm KS}$ can be reduced
by randomly redistributing stars within this range of $V_{\rm LSR}$. The
dotted curve in \figref{fig:KS} shows the dependence of $P_{\rm KS}$ on
$\Delta V$ when the randomised sample is compared to the model. The peak in
$P_{\rm KS}$ rises by three orders of magnitude and shifts from $\Delta
V=6.9\kms$ to $5.8\kms$.

Thus the stars that should be least affected by spiral structure suggest that
DB98 underestimated $V_\odot$ by $\Delta V=5.8\kms$, so the true solar motion
is $V_\odot=11\kms$. The systematic error $\sim1\kms$ on this value is
clearly much greater than the formal error reported by DB98.  In
\figref{fig:solarm} the black rectangle marks this revised solar motion.

It is interesting to test the extent to which Stromberg's equation is
verified by pseudo-isothermal \df s for main-sequence stars of a given
colour. The dashed curve in \figref{fig:solarm} shows the relation
between $S^2\simeq(1.2\sigma_R)^2$ (DB98) and $V_{\rm a}$ that one obtains by
calculating these quantities for a \df\ of the form (\ref{thinDF}) with
$\tau_{\rm m}$ increasing from $1$ to $11\Gyr$; over this age range
$\sigma_R$ increases from $12.4$ to $28\kms$. The dashed curve is plotted on
the assumption that the solar motion is $V_\odot=11\kms$, as marked by the
black rectangle. At low $S^2$ the slope of the dashed curve is similar to the
slope of the observational relation, but the slope flattens perceptibly with
increasing $S^2$. This flattening implies that the square bracket in 
Stromberg equation (\ref{eq:Stromberg}) diminishes with increasing velocity
dispersion. The is no evident reason why this bracket should be constant.

It is not inconceivable that spiral structure and/or the
bar have shifted the observational points with
$S^2$ in the range $(250,700)$  downwards from a relation
that runs from the black square, between the red points and on to just above
the points at $S^2>1200$. Moreover, a proponent of the conventional value of
$V_\odot$ should worry that if the dashed curve were moved down to start at
that value of $V_\odot$, it would lie below nearly all the data points. We
conclude that although we cannot confidently recommend an upward revision of
$V_\odot$, considerable caution should be exercised in the use of the
conventional value and more work is needed on the effect that spiral
structure has on the local velocity space.

Analysis of the space velocities of 18 maser sources for which trigonometric
parallaxes are available \citep{Reid09} provides tentative support for
$V_\odot$ being revised upwards to $11\kms$ \citep{McMillanB09}.

The values of $V_\odot$, the proper motion of Sgr A$^*$,
$6.38\pm0.04\,\hbox{mas\,s}^{-1}$ \citep{ReidB04}, and the distance to Sgr,
A$^*$, $8.33\pm0.31\kpc$ \citep{Gillessen09},   determine the local
circular speed $\vc(R_0)=R_0\mu_{A*}-V_\odot=(251\pm12-V_\odot)\kms$.
\cite{Flynn06} estimate the absolute I-band luminosity of the Galaxy to be
$M_I=-22.3$. At this absolute magnitude the ridge-line of the I-band
Tully--Fisher relation \citep{Dale99} gives a circular speed of only
$190\kms$; $251\kms$ lies $2.6\sigma$ from the ridge line. Thus the likelihood of
the Galaxy in the context of the Tully--Fisher relation is small but
increases rapidly with $V_\odot$, and this fact provides further support for
an upward revision of $V_\odot$.

\begin{figure}
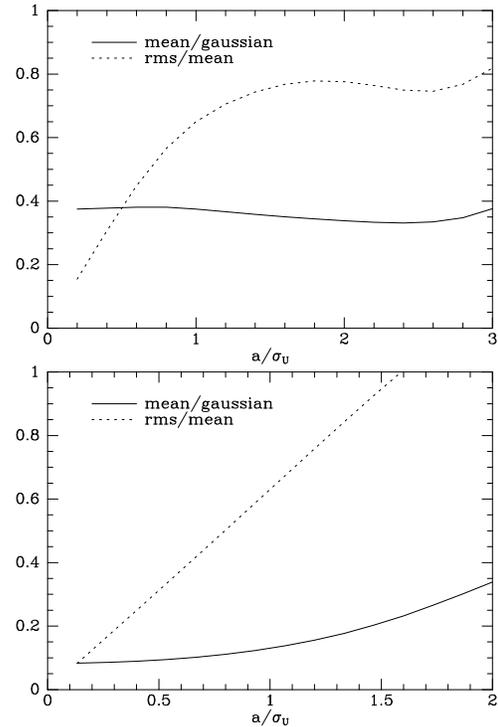

\centerline{\epsfig{file=bensby0,width=.8\hsize}}
\centerline{\epsfig{file=bensby1,width=.8\hsize}}
\caption{Full curves: the  mean value of the thin-disc \df\ (upper panel) and the
thick-disc \df\ (lower panel) over an ellipsoid in velocity space on which the
model \df\ (\ref{Bensbyeq}) is constant, divided by that constant value.
Dotted curves: the rms variation over an ellipsoid in the model \df\
divided by  the mean
value on the ellipsoid. The $x$ axis gives the semi-major axis of the
ellipsoid in multiples of $\sigma_U$.}\label{fig:bensby}
\end{figure}
 
\subsection{Distinguishing the thin and thick discs}

Studies of the chemistry of the thick disc depend heavily on identifying
nearby, bright stars that belong to the thick disc as targets of
medium-dispersion spectroscopy
\citep{Fuhrmann98,Bensby03,Venn04,Bensby05,Gilli06,Reddy06}. A popular
strategy for identifying target stars is to assume ellipsoidal velocity
distributions for each disc of the form \citep{Bensby03}
 \[\label{Bensbyeq}
f(U,V,W)\propto\exp\left(-{U^2\over2\sigma_U^2}-{(V-V_{\rm
a})^2\over2\sigma_V^2}-{W^2\over2\sigma_W^2}\right),
\] 
 where $(U,V,W)$ are velocity components with respect to the LSR, and each
component is assigned assumed values of $\sigma_U,\sigma_V,\sigma_W$ and the
rotational lag $V_{\rm a}$. A given star is assigned to the population for
which it gives the largest value of the \df\ that follows from equation
(\ref{Bensbyeq}) and 
assumed fractions of local stars that belong to each population.

\begin{figure*}
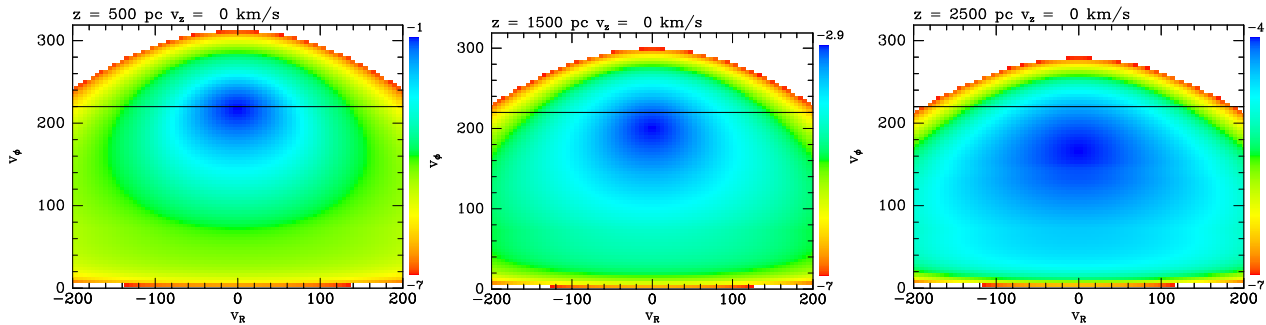

\centerline{\epsfig{figure=s500.ps,width=.31\hsize}\ 
\epsfig{figure=s1500.ps,width=.31\hsize}\ 
\epsfig{figure=s2500.ps,width=.31\hsize}}
\caption{Distributions in the $(v_R,v_\phi)$ plane for $v_z=0$ at three
distances from the plane: $0.5\kpc$ (left), $1.5\kpc$ (centre) and $2.5\kpc$.
Exactly these distributions are predicted at $z=0$ and $v_z=30$, $67$ and
$97\kms$.}\label{fig:corresp}
\end{figure*}

The idea behind equation (\ref{Bensbyeq}) is that, for appropriate
parameters, $f$ provides a useful approximation to the \df s of the two
discs. Here we investigate the quality of this approximation by comparing $f$
with the thin- and thick-disc components of the standard \df\ when the
parameters in $f$ take the values used by \cite{Bensby03}, which are given in
Table~\ref{tab:params}.  \figref{fig:bensby} quantifies the quality of the
approximation provided by $f$ by reporting the fractional \rms\ variation in
the appropriate component of the standard \df\ over each ellipsoidal surface
in velocity space on which $f$ is constant (dotted curves) -- ideally this
would vanish.  The full curves show the mean value of the \df\ over an
ellipsoid of constant $f$, divided by the value of $f$ on that ellipsoid --
since the normalisation of $f$ is arbitrary, the full lines in
\figref{fig:bensby} may be shifted up or down at will, but must be horizontal
if $f$ it to provide a useful approximation to the relevant component of the
standard \df.  

The upper panel of \figref{fig:bensby} is for the thin disc and the
lower panel for the thick disc.  Since the full curve in the upper panel is
approximately horizontal, we conclude that $f$ decreases from small to large
ellipsoids in the same way that our model thin-disc \df\ does. However, from
the fact that the dotted curve lies above $0.7$ for $a>\sigma_U$, we conclude
that the model \df\ varies by of order itself over the larger velocity
ellipsoids of the thin disc. Thus $f$ is a useful but not very accurate
approximation to the model \df\ for the thin disc.

The lower panel of \figref{fig:bensby} implies that $f$ provides a very poor
approximation to our model thick-disc \df: not only does the full curve rise
by more than a factor of 4 as $a$ increases to $2\sigma_U$, but the essentially linear
rise of the dotted curve implies that the model \df\ varies strongly over
ellipsoids of constant $f$.
What prevents $f$ providing a good approximation to the \df\ is the
continuous increase in the asymmetric drift as the numerical value of the
\df\ decreases. This increase is apparent in the downward motion of
isodensity contours in \figref{fig:UVplane} and drives the fall in $\langle
v_\phi\rangle$ with $z$ in the bottom right panel of \figref{fig:thinthick}.
In particular, at $z=0$ the density of stars in velocity space peaks close to
the circular speed in the thick disc as in the thin. This fact conflicts with
the structure of equation (\ref{Bensbyeq}).

\subsection{Local and in-situ samples}

The standard \df\ predicts that the distribution of stars in the
$(v_R,v_\phi)$ plane at height $z$ and $v_z=0$ is identical to the
$(v_R,v_\phi)$ distribution of stars at any other height $z'<z$ and velocity
$v_z'=\{2[\Phi_z(z)-\Phi_z(z')]\}^{1/2}$. Moreover, a sample of stars observed at some
distance $z$ from the plane will be heavily weighted towards stars whose
vertical motions have turning points there. So there should be a close
correspondence between local stars with a particular value of $v_z$ and
samples of stars at a given value of $z$. Since the SDSS
and its successors provide photometric distances and
proper motions  for millions of stars that lie $\gta1\kpc$
from the plane, and the GCS catalogue provides space velocities for over ten
thousand nearby stars,  this correspondence can be tested in some detail.
\figref{fig:corresp} shows some sample $(v_R,v_\phi)$ distributions.

\section{Conclusions}

We have explored the ability of distribution functions to provide models of
the thin and thick discs of the Milky Way. Our \df s are analytic functions
of the actions of orbits, which ensures that there is an intuitive relation
between the observable properties of the population a \df\ describes and the
functional form of the \df, and a meaningful way to compare models that use
different gravitational potentials. In this paper we have used expressions
for the actions that are only approximate, and imply that a star's vertical
motion is adiabatically invariant during the star's motion parallel to the
plane. In a forthcoming paper (McMillan et al., in preparation) orbital tori
will be used to eliminate this approximation, and thus quantify its validity.

We have shown that the vertical density profile and kinematics of the disc
are accurately modelled by the extremely simple \df\ (\ref{powerdf}).
However, we rejected this \df\ because  it is essential to be able to
break the \df\ for the thin disc down at least into contributions from stars
of various ages, and ideally into contributions from ranges in both age and
metallicity. That is, we must recognise that the Galaxy is built up of
innumerable stellar populations of various ages and metallicities, and each
population has its own \df. 
In this paper we have only begun to explore the resulting complexity by ascribing a
single \df\ to the thick disc and modelling the thin disc as a superposition
of \df s for stars of different ages. In reality both  discs are chemically
inhomogeneous and we should assign a distinct \df\ to the stars born at each
time with each chemical composition \citep[e.g.][]{SchoenrichBII}.
Hence the \df s presented in this paper should be considered
building blocks  from which more elaborate \df s may be in due course
constructed.

Our most basic building block is a ``pseudo-isothermal'' population of stars.
\figref{fig:HF} shows that the vertical distributions of young stellar
populations is well modelled by a pseudo-isothermal population. The density of a
pseudo-isothermal population does not decline exponentially with $z$, but
Figs.~\ref{fig:thinD} and \ref{fig:thinthick} show that, remarkably, the
composite population produced by stochastic acceleration of stars does have
an exponentially decreasing density profile. An excellent fit to the observed
density profile of the entire disc is obtained when a pseudo-isothermal thick disc
is added to the composite thin disc. The dispersion in $v_z$ of thin-disc
stars increases from $17.4\kms$ in the plane to $33\kms$ at $2.5\kpc$, while
that of the thick-disc stars increases from $\sim 35\kms$ in the plane to
$\sim48\kms$ at $2.5\kpc$. The thick disc contributes to the solar cylinder
24 per cent of the luminosity contributed by the thin disc, or 19.4 per cent
of the total luminosity of the disc.

Even though we are assuming that the dynamical coupling between motions in
and perpendicular to the plane is weak, two features of our \df s lead to
strong correlations between distributions in $v_R$ and $v_z$. One feature is
the fact that random velocities must increase as one moves inwards, and the
other is the simultaneous increases in $\sigma_R$ and $\sigma_z$ that are
driven by stochastic acceleration of a coeval population. Comparison of
Figs.~\ref{fig:DF} and \ref{fig:thinthick} show that, on account of this
correlation, the distribution of local stars in the $(v_R,v_\phi)$ plane is
atypical of the stellar population of the whole solar cylinder in just such a
way that our composite disc \df\ can simultaneously provide reasonable
matches to the very different
shapes of the distributions of GCS stars in $v_R$ and $v_\phi$. The widths of
the model distributions in $v_R$ and $v_\phi$ are controlled by a single
parameter, $\sigma_{r0}$. The shape of the $v_R$ distribution is
predetermined by our choice of the \df s functional form. The value of the
parameter $R_\d$ provides limited control of the shape of the $v_\phi$
distribution and we obtain the best fit to the observed distribution when
this parameter is chosen such that the disc's surface density declines
roughly exponentially with scale length $2.5\kpc$, which happens to agree
with the scale length inferred from near-IR star counts by \cite{Robin03}.

In principle the \df\ of the thick disc should be tightly constrained by the
dependence on  $z$ of the velocity dispersions $\sigma_R$ and $\sigma_z$.
These dependencies have recently been determined for SDSS stars by two
groups. Unfortunately, their results seem to be incompatible and the reasons
for the conflict are unknown. 

The standard model provides an excellent fit to the seminal work of Kuijken
\& Gilmore, perhaps because the gravitational potential in which the \df\ is
evaluated was partly fitted to that work. Some of the difficulties
encountered here with fitting newer data may arise from inaccuracy of the
potential used.  A worthwhile exercise would be to fit data from the GCS,
RAVE and SDSS surveys to models that combined \df s of the type presented
here with and a multi-parameter gravitational potential: by simultaneously
fitting the parameters in both the \df\ and the potential, one should be able
to obtain reasonable fits to the data, providing the latter have been purged
of such evident inconsistencies as those seen in \figref{fig:powersig}. Data
from more than one survey would probably have to be used since SDSS stars are
too faint to constrain the thin disc tightly, although the RAVE survey, which
certainly probes the thick disc effectively, may include enough nearby stars
to make the Hipparcos-based GCS survey obsolete.

The model fit to the $v_\phi$ distribution of GCS stars is far from perfect.
Some of the disagreement arises because, as is well known, the Galactic bar
and spiral arms give rise to features (``star streams'') in the local
velocity distribution that are inconsistent with the Galaxy being
axisymmetric and in a steady state, as our models assume.  Our favoured model
$v_\phi$ distribution would fit the data better if the conventional value of
the solar motion $V_\odot$ were $\sim6\kms$ too low. Tentative support for
such an increase in the $V_\odot$ is provided by astrometry of stellar masers
\citep{Reid09,McMillanB09}, and any increase would also tend to bring the
Galaxy more into line with the Tully--Fisher relation between $\vc$ and $M_I$
for external galaxies.  By systematically perturbing the velocities of all
solar-neighbourhood stars, spiral structure might lead to the classical
approach to the determination of $V_\odot$ yielding an underestimate. Further
work is required to explore this possibility, and at this stage we would
merely stress that the systematic error in $V_\odot$ is much larger than the
formal errors given by DB98 and \cite{AumerB}.

In the models, the asymmetric drifts of both the thin and thick discs increase
with height. A disc's asymmetric drift is largely controlled by its parameter
$R_\d$ and in the standard model the asymmetric drift of the thin disc
exceeds that of the thick disc above $1\kpc$ because we have adopted a
slightly larger value of $R_\d$ for the thick disc than for the thin disc.

A popular strategy for assigning solar-neighbourhood stars to the thin or
thick disc is to find the values taken by each disc's model \df\ at the
star's location. The model \df s used are perfectly ellipsoidal but we show
that such \df s provide poor approximations to the thick-disc component of
the standard \df, so a markedly cleaner separation of the two discs could be
obtained by replacing the ellipsoidal \df s by the thin- and thick-disc
components of the  standard \df.

Although the observational material relating to the Galaxy has increased
enormously in recent years, we have shown that much of the available data can
be successfully modelled with a simple analytical \df. In a couple of aspects
the data are in mild conflict with the \df, but it is at least as likely
that the fault lies with the data as the \df. In the coming decade the
volume and quality of the observational material available will increase
dramatically. We anticipate that comparisons between each new data set and an
evolving standard \df\ will reveal successes and failures similar to those
encountered here. The successes will confirm the value of the \df\ as a
summary of a large and inhomogeneous body of data, and the failures will lead
to critical re-examination of both data and \df. Sometimes the failure will
arise from a defective calibration of the data or incorrect assumptions
used in its reduction, and other times it will indicate that the \df\ is too
simplistic.  Either way we will learn something new and interesting.

In this paper the \df's parameters have been fitted to the data by eye and no
attempt has been made to quantify uncertainties in parameter values. Clearly
such uncertainties are important, and they could be most securely established
by carrying the \df's predictions closer to the raw observations than we have
done.  In future work probability distributions in
colour--magnitude--proper-motion space, etc., should be predicted that can be
compared with the actual star counts.

Upcoming infrared surveys, such as the VHS with Vista and APOGEE, will probe
the disc at remote locations. The predictions of the standard \df\ for those
locations will be presented shortly, after orbital tori have been introduced
as the means to convert between Cartesian and angle-action variables. This
upgrade will make obsolete the approximation of adiabatically
invariant vertical motions  used here.

\section*{Acknowledgements}
I thank Michael Aumer for providing the data shown in \figref{fig:solarm} and
Zeljko Ivezic for providing the data plotted in \figref{fig:ivezic}.
The members of the Oxford dynamics group contributed valuable comments on drafts of
this work. It is a pleasure to acknowledge valuable conversations with Ralph
Sch\"onrich.

\label{lastpage}
\end{document}